\documentstyle[twocolumn,prl,aps]{revtex}
\begin{document}
\draft
\title{Isospin Dependence of the Spin Orbit Force and
Effective Nuclear Potentials}
\author{M.M. Sharma$^1$, G. Lalazissis$^2$, J. K\"onig$^2$,
and P. Ring$^2$}
\address{$^1$Max Planck Institut f\"ur Astrophysik,
Karl-Schwarzschildstrasse 1, D-85740 Garching, Germany}
\address{$^2$Physik-Department der Technischen
Universit\"at M\"unchen, D-85747 Garching, Germany}
\maketitle
\begin{abstract}
The isospin dependence of the spin-orbit potential is
investigated for an effective Skyrme-like energy functional
suitable for density dependent Hartree-Fock calculations.
The magnitude of the isospin dependence is obtained from a
fit to experimental data on finite spherical nuclei. It is
found to be close to that of relativistic Hartree models.
Consequently, the anomalous kink in the isotope shifts of
Pb nuclei is well reproduced.
\end{abstract}
\pacs{PACS numbers : 21.60Cs, 21.10.Dr, 21.10.Gv}

The Hartree-Fock approach based upon phenomenological
density dependent forces\cite{VB.72,FQ.78,DG.80} has proved
to be very successful in the microscopic description of
ground state properties of nuclear matter and of finite
nuclei over the entire periodic table. In all these
calculations the spin-orbit potential has been assumed to
be isospin independent. Its only parameter, the strength,
is usually adjusted to the experimental spin-orbit
splitting in spherical nuclei like $^{16}$O or $^{208}$Pb.
The exchange term, however, causes for nuclei with neutron
excess a strong isospin dependence of the corresponding
single-particle spin-orbit field.

In recent years Relativistic Mean Field (RMF)
theory\cite{SW.86} with nonlinear self-interactions between
the mesons has gained considerable interest for the
investigations of low-energy phenomena in nuclear
structure. With only a few phenomenological parameters such
theories are able to give a quantitative description of
ground state properties of spherical and deformed
nuclei\cite{GRT.90,SNR.93} at and away from the stability
line. In addition, excellent agreement with experimental
data has been found recently also for collective
excitations such as giant resonances\cite{VBR.94} and for
twin bands in rotating superdeformed nuclei\cite{KR.93}. In
many respects the relativistic mean-field theory is
regarded as similar to the density-dependent Hartree-Fock
theory of the Skyrme type\cite{Thi.86,Rei.89}

Recently, however, detailed investigations of high
precision data on nuclear charge radii in Pb
isotopes\cite{TBF.93,SLR.93} and of shell effects at the
neutron drip line\cite{Dob.94,SLH.94} have shown
considerable differences between the Skyrme approach and
the relativistic mean field theory. In fact, density
dependent Hartree-Fock calculations with
Skyrme\cite{TBF.93} or Gogny forces\cite{Egi.94}, which
used so far antisymmetrized, isospin independent spin-orbit
interactions, were not able to reproduce the kink in the
isotope shifts of Pb nuclei (see Fig. \ref{F1}). On the
other hand, this kink is obtained in the RMF theory without
any new adjustment of parameters\cite{SLR.93}. Another
considerable difference has been found in theoretical
investigation of shell effects in very exotic Zr-isotopes
near the neutron drip line: in conventional
non-relativistic Skyrme calculations the shell gap at
isotope $^{122}$Zr with the magic neutron configuration N =
82 is totally smeared out\cite{Dob.94}, whereas
relativistic calculations using various parameter sets show
at N=82 a clear kink in the binding energy as a function of
the neutron number\cite{SLH.94}.  This difference is caused
by the different spin-orbit splitting of the single
particle levels in these nuclei. Mass calculations within
the Finite Range Droplet Model (FRDM) \cite{MNM.94}, which
are based on an isospin dependent spin-orbit term carefully
adjusted to experimental data, are in excellent agreement
with the relativistic predictions.

This gives us a hint, that one does not really need full
relativistic calculations in order to understand these
differences between conventional density dependent
Hartree-Fock calculations and RMF theory. A spin-orbit term
with a properly chosen isospin dependence might represent
the essential part of a relativistic calculation. In this
letter, we therefore explore the isospin dependence of the
spin-orbit term in non-relativistic Skyrme calculations and
analyze the consequences of this on the nuclear properties.
We start from the Skyrme type force
\begin{eqnarray}
V(1,2)&=&t_0(1+x_0P^\sigma)
\delta({\bf r}_1-{\bf r}_2)\nonumber\\
&+&~t_1(1+x_1P^\sigma)
(\delta({\bf r}_1-{\bf r}_2){\bf k}^2+h.c.)\nonumber\\
&+&~t_2(1+x_2P^\sigma){\bf k}\delta({\bf r}_1-{\bf r}_2){\bf k}
\label{mska}\\
&+&~\frac{1}{6}t_3(1+x_3P^\sigma)
\delta({\bf r}_1-{\bf r}_2)\rho^\alpha
\nonumber\\
&+&~W_0(1+x_wP^\tau)(
{\mbox{\boldmath $\sigma$}}^{(1)}+
{\mbox{\boldmath $\sigma$}}^{(2)})
{\bf k}\times\delta({\bf r}_1-{\bf r}_2){\bf k}
\nonumber
\end{eqnarray}
with ${\bf k}=\frac{1}{2}({\bf p}_1-{\bf p}_2)$. In
contrast to the conventional Skyrme ansatz, where the
energy functional contains a Hartree- and a
Fock-contribution, we here neglect the exchange (Fock-)
term for the spin-orbit potential in the last line of Eq.
(\ref{mska}).  Otherwise, the operator $P^\tau$ would be
equivalent to $+1$ for spin saturated systems. For the rest
of the potential the Fock terms are included. The
associated 11 parameters $t_i$, $x_i$ ($i=0\dots 3$),
$W_0$, $x_w$ and $\alpha$ of the Modified Skyrme Ansatz
(MSkA) in Eq. (\ref{mska}) are determined by a fit to
experimental data of finite spherical nuclei. The nuclear
properties taken into consideration are the empirical
binding energies and charge radii of the closed-shell
nuclei $^{16}$O, $^{40}$Ca, $^{90}$Zr, and $^{208}$Pb. In
order to take into account the variation in isospin we have
also included Sn-isotopes $^{116}$Sn, $^{124}$Sn, and the
doubly closed nucleus $^{132}$Sn as well as one of the lead
isotopes $^{214}$Pb. The resulting force and its parameters
are presented in Table \ref{T1}.

The spin-orbit term in the single-particle field derived
>from this force has the form ${\bf W}_\tau({\bf r})( {\bf
p}\times{\mbox{\boldmath $\sigma$}})$ with
\begin{equation}
{\bf W}_\tau({\bf r})~=~
W_1{\mbox{\boldmath $\nabla$}}\rho_\tau+,
W_2{\mbox{\boldmath $\nabla$}}\rho_{\tau^\prime\ne\tau},
\label{spinorbit}
\end{equation}
where $\rho_\tau$ is the density for neutrons or protons
($\tau=n$ or $p$) and $W_1 =W_0(1+x_w)/2$, $W_2 =W_0/2$.
Conventional Skyrme calculations use a spin-orbit potential
without isospin dependence and include the Fock term. This
leads to $x_w=1$ and the relationship $W_1/W_2=2$, which is
nearly by a factor 2 different from the value $1.0005$
obtained within the modified Skyrme Ansatz MSkA (see Table
\ref{T1}). It is interesting to note that the fit leads to
a value of $x_w$ very close to zero, which corresponds to
the one without the Fock term.

In order to study the spin-orbit term in the RMF theory we
start from the standard Lagrangian density\cite{GRT.90}
\begin{eqnarray}
{\cal L}&=&\bar\psi\left(\gamma(i\partial-g_\omega\omega
-g_\rho\vec\rho\vec\tau-eA)-m-g_\sigma\sigma
\right)\psi
\nonumber\\
&&+\frac{1}{2}(\partial\sigma)^2-U(\sigma )
-\frac{1}{4}\Omega_{\mu\nu}\Omega^{\mu\nu}
+\frac{1}{2}m^2_\omega\omega^2\nonumber\\
&&-\frac{1}{4}{\vec{\rm R}}_{\mu\nu}{\vec{\rm R}}^{\mu\nu}
+\frac{1}{2}m^2_\rho\vec\rho^{\,2}
-\frac{1}{4}{\rm F}_{\mu\nu}{\rm F}^{\mu\nu}
\label{lagrangian}
\end{eqnarray}
which contains nucleons $\psi$ with mass $m$.  $\sigma$-,
$\omega$-, $\rho$-mesons, the electromagnetic field and
nonlinear self-interactions $U(\sigma)$ of the
$\sigma$-field,
\begin{equation}
U(\sigma)~=~\frac{1}{2}m^2_\sigma\sigma^2+
\frac{1}{3}g_2\sigma^3+\frac{1}{4}g_3\sigma^4.
\end{equation}

In a non-relativistic approximation of the corresponding
Dirac equation\cite{Koe.94} we obtain the following single
particle spin-orbit term:
\begin{eqnarray}
{\bf W}_\tau({\bf r})&=&
\frac{1}{m^2m^{*2}}(C^2_\sigma+C^2_\omega+C^2_\rho)
{\mbox{\boldmath $\nabla$}}\rho_\tau\nonumber\\
&&~+\frac{1}{m^2m^{*2}}(C^2_\sigma+C^2_\omega-C^2_\rho)
{\mbox{\boldmath $\nabla$}}\rho_{\tau^\prime\ne\tau},
\label{relspinorbit}
\end{eqnarray}
with $C_i^2=(m\,g_i/m_i)^2$ for $i=\sigma,\omega,\rho$,
which is similar in form to the spin-orbit field
(\ref{spinorbit}), but which contains $\bf r$-dependent
parameters
\begin{eqnarray}
W_1&=&\frac{1}{m^2m^{*2}}(C^2_\sigma+C^2_\omega+C^2_\rho)\\
W_2&=&\frac{1}{m^2m^{*2}}(C^2_\sigma+C^2_\omega-C^2_\rho),
\end{eqnarray}
with $m^*({\bf r})=m-g_\sigma\sigma({\bf r})$
The $r$-dependence drops out in the ratio $W_1/W_2$ which
is $1.13$ for the parameter set NL1\cite{GRT.90} and $1.10$
for the parameter set NL-SH\cite{SNR.93}.  This is only
slightly higher than the value $1.0005$ obtained in the
Modified Skyrme Ansatz by the fit to the empirical data
(see Table \ref{T1}). The absolute size of the spin-orbit
term turns out to be $\bf r$-dependent, which stems from
the $\bf r$-dependence of the effective mass
$m^*({\bf r})$. It can be approximated by $m/m^*\approx
1+C^2_\sigma/m^3\,\rho({\bf r})$. A more careful
consideration would therefore require an explicitly density
dependent spin-orbit term.  It has not been included in the
present investigation.

In Table \ref{T2} we show nuclear matter results obtained
in the MSkA and compare it with the values from the
conventional Skyrme force SkM$^*$\cite{BQB.82}. The
saturation density is obtained as $\rho_0=0.1531$ fm$^{-3}$
in MSkA.  This is the same as the value obtained from an
extensive fit of the mass formula FRDM\cite{MNM.94}. The
binding energy per particle $E/A$ is 16.006 MeV, which is
close to that of other Skyrme forces. The compression
modulus $K=319$ MeV is somewhat higher than that of the
presently adopted Skyrme forces, but it lies within
the error bars of the analysis based upon the breathing
mode energies. The asymmetry energy $J$ is close to the
empirical value of 33 MeV. The effective mass $m^*$ is in
good agreement with that of SkM$^*$.

In Table \ref{T3} we show binding energies and charge radii
obtained in the MSkA for a number of spherical nuclei and
compare them with those from SkM$^*$.  A comparison with
the empirical values shows, that the binding energies
obtained with MSkA have improved over those of SkM$^*$.
The slightly reduced charge radii in MSkA seem to be
connected with a slightly higher binding energy, which
improves the results for the lighter Pb-isotopes. In
general the charge radii are improved including that of
$^{16}$O.

In Fig. 1 we show the isotope shifts of Pb nuclei for the
Modified Skyrme Ansatz MSkA together with experimental data
and results of conventional Skyrme calculations. For MSkA
we observe a clear kink at the double magic nucleus
$^{208}$Pb, whereas the conventional Skyrme force SkM$^*$
with an isospin independent spin-orbit term gives an almost
straight line. For the lighter isotopes both theories give
excellent agreement, on the heavier side MSkA comes closer
to the experimental isotope shifts.  It may be recalled,
that the RMF theory\cite{SLR.93} is successful in
reproducing the full size of this kink. MSkA uses an
isospin dependent, but density independent spin-orbit
force.  In contrast the spin-orbit term derived from the
RMF theory (see Eq. \ref{relspinorbit}) is implicitly
density dependent through the density dependence of the
effective mass $m^*({\bf r})$.  A density dependence of the
spin-orbit term in Skyrme theory might improve the charge
radii of the heavier Pb-isotopes. It has also been observed
that a density dependent pairing force can possibly improve
the situation in this context\cite{TBF.93}. This requires,
however, further investigations.

In Fig. 2 we present binding energies of Zr-nuclei about
the neutron drip line as a function of the mass number. It
is observed that in agreement with earlier
investigations\cite{Dob.94} within the conventional Skyrme
theory shell effects about the closed shell nucleus
$^{122}$Zr are weakened considerably. In contrast,
RMF-calculations exhibit strong shell effects in the drip
line region. It has been surmised in Ref. \cite{SLH.94}
that this is caused by the differences in the spin-orbit
terms in the two approaches. It is gratifying to see that
introduction of an isospin dependent (not antisymmetrized)
spin-orbit term in Skyrme theory leads to stronger shell
effects. This is also in agreement with the predictions of
the FRDM\cite{MNM.94}, where also an isospin dependent
spin-orbit term is used. A more quantitative analysis shows
that the ratio $W_1/W_2$ in FRDM is close to 1.06 for the
Pb-nuclei and 1.09 for $^{122}$Zr.

Summarizing, we conclude that the isospin dependence of the
spin-orbit term has an essential influence on the details
of anomalous isotope shifts of Pb-nuclei. A new Modified
Skyrme Ansatz has been proposed, and the isospin dependence
of the spin-orbit strength has been determined. The
magnitude of this isospin dependence $x_w$ is in agreement
with the deductions from the relativistic mean-field
theory. A reasonably good agreement with the experimental
data on the binding energies and charge radii has been
obtained. The kink in the isotope shifts of Pb-nuclei has
been obtained in the modified Skyrme ansatz.  However, the
agreement with the empirical isotope shifts for heavy
Pb-nuclei is not so good as in the RMF theory. This calls
for further investigations including a density dependence
of the spin-orbit term.

Finally a remark about the isospin dependence of the
spin-orbit term: our investigations lead to the interesting
result, that the parameter $x_w$ in the Eq. (1) is close to
zero. This leads practically to an isospin independent
single-particle spin-orbit field in Eq. (2). This is in
agreement with relativistic calculations, where the entire
isospin-dependence of the spin-orbit field is caused by the
parameter $C_\rho$ in Eq. (5), which is in fact rather
small. In contrast the two-body spin-orbit potential in
conventional Skyrme theory is isospin independent. The
exchange term, however, causes a strong isospin dependence
in the corresponding single-particle spin-orbit field. On
the other hand in RMF theory the spin-orbit field has its
origin in Lorentz covariance. There is no contribution from
a two-body spin-orbit potential and an exchange term is
therefore excluded.

One of the authors (G.A.L) acknowledges support from the
E.U., HCM program, contract: EG/ERB CHBICT-930651 , This
work is also supported in part by the Bundesministerium
f\"ur Forschung und Technologie under the project 06 TM
743.

\begin{figure}
\caption{Isotopic shifts in the charge radii of Pb isotopes normalized
to the nucleus $^{208}$Pb as a function of the mass number A in the
Modified Skyrme Ansatz (MSkA) as compared to the Skyrme force SkM$^*$
and the empirical data\protect{\cite{Ott.89}}.}
\label{F1}\end{figure}

\begin{figure}
\caption{Binding energy of Zr isotopes at neutron drip line in the
MSkA and SkM$^*$ calculation.}
\label{F2}
\end{figure}

\begin{table}
\caption{Parameters of the interaction in the modified Skyrme
Ansatz (MSkA).}
\begin{tabular}{ll}
  $t_0 = -1200.074$ (MeV fm$^3$) & $x_0 = 0.187$ \\
  $t_1 = 396.302$ (MeV fm$^5$)   & $x_1 = 0.018$ \\
  $t_2 = -105.579$ (MeV fm$^5$)  & $x_2 = -0.059$\\
  $t_3 = 10631.527$ (MeV fm$^{3+3\alpha}$) & $x_3 =  0.046$ \\
  $W_0 = 316.38$ (MeV fm$^5$) & $x_w = 0.0005 $   \\
  $\alpha = 0.7557 $  & \\
\end{tabular}
\label{T1}
\end{table}

\begin{table}
\caption{Nuclear matter properties obtained in the modified Skyrme
Ansatz (MSkA).}
\begin{tabular}{lll}
&   MSkA & SkM$^*$\\
\hline
 $\rho_0$       & 0.1531 fm$^{-3}$ & 0.1603 fm$^{-3}$ \\
 $(E/A)_\infty$ & 16.006 MeV       & 15.776 MeV       \\
 $K$            & 319.4  MeV       & 216.7  MeV       \\
 $\rho_0^2e'''$ & 100.2 MeV        &  57.9  MeV       \\
 $J$            & 30.0 MeV         &  30.0  MeV       \\
 $m^*/m$        & 0.76             &   0.79           \\
\end{tabular}
\label{T2}
\end{table}

\begin{table}
\caption{The binding energies and charge radii obtained with the
Modified Skyrme Ansatz (MSkA) in the Hartree-Fock approximation as
compared with the normal Skyrme force SkM* and the empirical values.}
\begin{tabular}{cccccccc}
 & \multicolumn{3}{c}{Binding Energies (MeV)}&
  & \multicolumn{3}{c}{ Charge Radii (fm)} \\
\cline{2-4} \cline{6-8}
Nuclei & expt.& MSkA &  SkM* && expt. & MSkA & SkM* \\
\hline
$^{16}$O&-127.6&-128.1& -127.7  && 2.730 & 2.742 & 2.811 \\
$^{40}$Ca&-342.1&-342.8& -341.1 && 3.450 & 3.468 & 3.518 \\
$^{48}$Ca&-416.0&-416.8& -420.1 && 3.500 & 3.506 & 3.537 \\
$^{90}$Zr&-783.9&-781.9& -783.0 &&4.270 & 4.274 & 4.296  \\
$^{116}$Sn&-988.7&-984.0& -983.4 && 4.626& 4.623 & 4.619 \\
$^{124}$Sn&-1050.0&-1047.9& -1049.0 && 4.673& 4.677 & 4.678 \\
$^{132}$Sn&-1102.9&-1106.3& -1110.7 &&   -   & 4.728 & 4.727 \\
$^{200}$Pb&-1576.4&-1570.9& -1568.4 && 5.464 & 5.465 & 5.468 \\
$^{202}$Pb&-1592.2&-1588.3& -1586.0 && 5.473 & 5.475 & 5.478 \\
$^{204}$Pb&-1607.5&-1605.4& -1603.4 && 5.483 & 5.486 & 5.489 \\
$^{206}$Pb&-1622.3&-1622.3& -1620.3 && 5.492 & 5.496 & 5.501 \\
$^{208}$Pb&-1636.5&-1637.8& -1636.4 && 5.503 & 5.506 & 5.510 \\
$^{210}$Pb&-1645.6&-1645.8& -1645.6 && 5.522 & 5.522 & 5.520 \\
$^{212}$Pb&-1654.5&-1653.8& -1654.5 && 5.540 & 5.537 & 5.531 \\
$^{214}$Pb&-1663.3&-1661.6& -1663.1 && 5.558 & 5.552 & 5.541 \\
\end{tabular}
\label{T3}
\end{table}
\end{document}